\documentclass[doublespacing]{elsart}
\usepackage[english]{babel}

\usepackage{epsfig}
\usepackage{graphics}
\usepackage{cite}

\usepackage{amssymb}

\begin{document}

\begin{frontmatter}

\title{Telegraph equations for the case of a waveguide with moving boundary}

\author{V.G. Baryshevsky}

\address{Research Institute for Nuclear Problems, Belarusian State
University, 11~Bobruiskaya Str., Minsk 220030, Belarus}
\ead{bar@inp.bsu.by, \\ v\_baryshevsky@yahoo.com}

\begin{abstract}
Telegraph equation describing the compression of electromagnetic
waves in a waveguide (resonator) with moving boundary are derived.
It is shown that the character of oscillations of the compressed
electromagnetic field depends on the parameters of the resonator,
and under certain conditions, the oscillations of voltage
(current) yield the exponential growth, leading to a noticeable
change in the radiation losses.
\end{abstract}

\end{frontmatter}
\section{Introduction}

 The experiments to produce ultrahigh magnetic fields using magnetic
 cumulative generators (flux compression generators), pioneered by A.D. Sakharov  \cite{Sakharov}  and C.M.
 Fowler \cite{Fowler}, gave rise to a rapidly developing research area related to high energy density physics (see, e.g.,
\cite{Neuber}).
A distinctive feature of these experiments is the use of
fast-moving under explosion shells (shock waves)  to compress the
magnetic field.
For example, in helical flux compression generators (FCGs), the
compressed   magnetic field proves trapped between the solenoid
and the liner. From the electrodynamical viewpoint, the closed
region containing a magnetic (an electromagnetic) field and
undergoing compression is a resonator with moving boundary.

The transformation of a microwave field during the  compression of
a cylindrical resonator was first studied by  L.P. Feoktistov and
V.V. Klimov  in  \cite{Feoktistov}. The authors of
\cite{Feoktistov} derived the explicit expression describing the
electromagnetic field inside the resonator in a particular case of
axisymmetric mode $TM_{010}$
  $(E_{001})$ of a compressed cylindrical resonator whose walls
move at constant velocity and showed that the energy of the field
in the resonator increases, because  the frequency of the field is
increased. Let us note here that the structure of FCG resonators
can be more complicated. The helical FCG, for example, is a
coaxial resonator whose mode structure differs considerably from
that of a cylindrical resonator. In particular, multiply-connected
cross section leads to the presence of a $TEM$ wave. The coaxial
resonator can be considered as a section of a coaxial waveguide
with closed boundaries, and telegraph equations  can be applied to
analyze oscillations in the resonator. However, commonly-known
telegraph equations \cite{Landau, Simonyi} are formulated for  the
case when the geometric dimensions of the line (coaxial waveguide)
are time-independent.

 In this paper, we derived  telegraph equations to describe
the  compression of a $TEM$ wave for the case of a waveguide
(resonator) with moving boundary. The derived equations are used
to analyze how the oscillations of the compressed electromagnetic
field depend on the parameters of the resonator.

\section{Telegraph equations for moving circuits}
Let us consider a circuit moving in the electromagnetic field with
a velocity   $\vec{v}(t)$. Due to relativistic effects, the
current induced in the conductor is \cite{Landau}
\begin{equation}
\label{n1} \vec{\jmath}=\sigma \left(\vec E+[\vec v \vec
B]\right),
\end{equation}
where $\vec E$ is the electric field strength and $\vec B$ is the
magnetic inductance. It follows from (\ref{n1}) that the
electromotive force  acting in a moving closed circuit is
\cite{Landau}
\begin{equation}
\label{n2} \varepsilon =\oint\limits_{l(t)} \left(\vec
E+\left[\vec v(t) \vec B\right]\right) d\vec l.
\end{equation}
According to Faraday's law, the electromotive force (\ref{n2})
equals the time rate of change of magnetic flux  $\Phi$:
\begin{equation}
\label{n3} \varepsilon = - \frac{d\Phi}{dt}.
\end{equation}
Taking the time derivative in (\ref{n1}), we should allow for the
total change of the magnetic flux through the circuit (associated
with both time change of the magnetic inductance and change in the
position of the  circuit)  \cite{Landau}.

Let us consider the section of length $\Delta x$ of the coaxial
waveguide and the circuit  $A\, B\, C\, D$ (see Fig. 1).

\begin{figure}[htbp]
\label{fig1}
\begin{center}
     \resizebox{40mm}{!}
       {\includegraphics{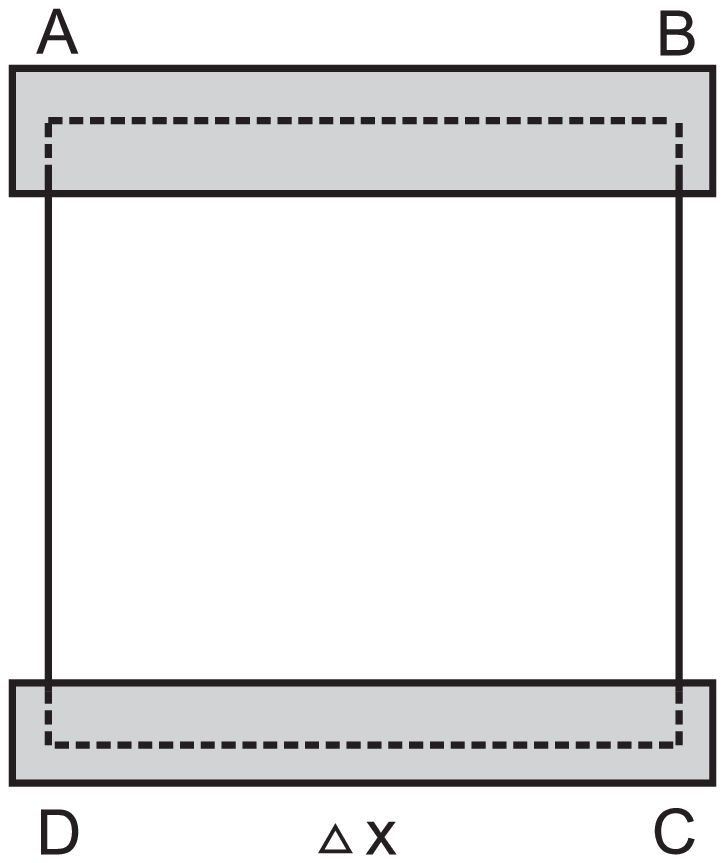}}\\
\caption{Section of length $\Delta x$ of the coaxial waveguide and
the circuit  $A\, B\, C\, D$.}
\end{center}
\end{figure}

 Let us introduce the inductance per unit length $L_0{(t)}$ and capacitance
 of the unit length  $C_0{(t)}$ of the line.
  Because the dimensions
 of the waveguide (the moving line) depend on time,
 $L_0$ and  $C_0$ also depend on time. The inductance and capacitance of
 the  waveguide section of length $\Delta x$ are $L_0
\Delta x$ and $C_0 \Delta x$,  respectively, and $I$ is the
current. The magnetic flux $\Phi$ through the circuit $A\, B\, C\,
D$ is
\[
\Phi=L_0 \Delta x I.
\]
This yields
\begin{equation}
\label{n4} L_0\frac{\partial I}{\partial t}\Delta x+
\frac{\partial L_0}{\partial t} I\Delta x + \varepsilon (x,\,
t)=0.
\end{equation}
Let us make use of the equality  \cite{Simonyi}
\begin{equation}
\label{n5}  \varepsilon (x,\, t)= R_0\Delta x I +\frac{\partial
U}{\partial x} \Delta x,
\end{equation}
where $R_0$ is the resistance of the unit length of the line and
 $U$ is the voltage between the conductors.

\begin{figure}[htbp]
\label{fig2}
\begin{center}
     \resizebox{70mm}{!}
       {\includegraphics{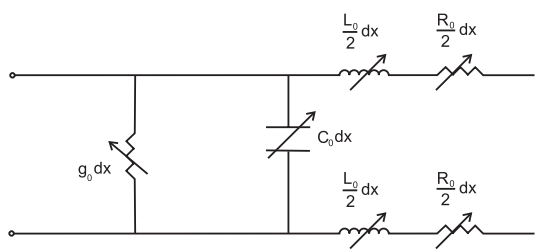}}\\
\caption{Circuit with time-dependent parameters.}
\end{center}
\end{figure}

As a result, we have
\begin{equation}
\label{n6}  L_0 (x,t)\frac{\partial I (x,t)}{\partial t}+
\frac{\partial L_0 (x, t)}{\partial t} I (x,t) + R_0 (x, t) I(x,
t) + \frac{\partial U (x,t)}{\partial x}= 0.
\end{equation}
The charge  $Q$ of the section having the length $\Delta x$ is
given by $Q=C_0\Delta x U (x,t)$. Using the charge density
continuity equation, we can obtain the following equality:
\begin{equation}
\label{n7} C_0\frac{\partial U}{\partial t} +\frac{\partial
C_0}{\partial t} U (x,t) + g_0 U +\frac{\partial U}{\partial x}=0.
\end{equation}
Here $g_0$, as in well as in the line with time-constant
capacitance,  describes   the leakage current. In a similar manner
as in the case of constant parameters of the system 
\cite{Simonyi}, we can present the system under consideration in
terms of the limit of the set of sequential elements (see Fig.
\ref{fig2}).

Let us write the derived pair of equations for  $I$ and $U$ in a
similar form as the standard telegraph equations with
time-independent parameters:
\begin{eqnarray}
\label{n8} - \frac{\partial U}{\partial x}=
L_0{(t,x)}\frac{\partial I}{\partial t}+ \left( \frac{\partial
L_0{(t, x)}}{\partial t}+ R_0 {(t, x)}\right) I
\nonumber\\
-\frac{\partial I}{\partial x}=C_0 {(t, x)}\frac{\partial
U}{\partial t}+\left ( \frac{\partial C_0{(t,x)}}{\partial t}+g_0
(t,x)\right) U.
\end{eqnarray}

This equation set together with the initial and boundary
conditions allows finding the current and voltage in a coaxial
resonator (waveguide, line) with moving conductors (with time and
coordinate-dependent  $L_0$, $C_0$, $R_0$, and  $g_0$).

The derivative  $\frac{\partial L_0}{\partial t}$ acts as
resistance (in the same manner as in the situation known in the
theory of FCG  \cite{Neuber}). The derivative  $\frac{\partial
C_0}{\partial t}$ has an effect on the leakage currents, which can
lead to the increase in the current instead of the drop in the
case when
 $\frac{\partial C_0}{\partial t}< 0$.

Now let us assume that the quantities   $L_0$, $C_0$, $R_0$, and
 $g_0$ are all independent of the coordinate  $x$. By differentiation
 of (\ref{n8}) with respect to the coordinate  $x$,
 we can obtain the following second-order equations for voltage and current:
 \begin{eqnarray}
 \label{n9}
 \frac{\partial^2 U}{\partial x^2} & = &
 L_0C_0 \frac{\partial^2
 U}{\partial t^2}+ \left[\frac{\partial L_0C_0}{\partial t}+
 L_0\frac{\partial C_0}{\partial t}+  L_0 g_0 + R_0 C_0\right]\frac{\partial
 U}{\partial t}\nonumber\\
 & + & \left[\frac{\partial}{\partial t}\left(L_0\frac{\partial
 C_0}{\partial t}\right) + \frac{\partial L_0 g_0 }{\partial t}+
 R_0 \frac{\partial C_0}{\partial t}+R_0 g_0 \right ] U,
 \end{eqnarray}

 \begin{eqnarray}
 \label{n10}
 \frac{\partial^2 I}{\partial x^2} & = &
 L_0C_0 \frac{\partial^2
 I}{\partial t^2}+ \left[\frac{\partial L_0C_0}{\partial t}+
 C_0\frac{\partial L_0}{\partial t}+ g_0 L_0  + R_0 C_0\right]\frac{\partial
 I}{\partial t}\nonumber\\
 & + & \left[\frac{\partial}{\partial t}\left(C_0\frac{\partial
 L_0}{\partial t}\right) + \frac{\partial C_0 R_0}{\partial t}  +
 g_0 \frac{\partial L_0}{\partial t}+g_0 R_0 \right ] I,
 \end{eqnarray}

We shall recall here that $L_0C_0=\frac{1}{v^2}$, where $v$ is the
wave's phase velocity. Equations (\ref{n9}) and (\ref{n10}) hold
true when   $L_0$, $C_0$, $R_0$, and
 $g_0$ are independent of  $x$. Otherwise we should apply  (\ref{n8}).

 Let us consider  the  coaxial resonator (waveguide section, line)
 with time-independent length  $l$ and varying
  transverse dimensions. Because the coefficients on the right-hand side
  of equations
   (\ref{n9}) and  (\ref{n10}) are independent of the coordinate  $x$, we can
   use the separation of variables method.
   As a result,  (\ref{n9}) yields

\begin{equation}
\label{n11} \frac{\partial^2 \varphi}{\partial x^2}+k^2 \varphi=0.
\end{equation}
With due account of the boundary conditions, this equation is
solvable for a set of eigenwavenumbers  $k_n$ and eigenfunctions
$\varphi_n(x)$ that are orthogonal and normalized.

By way of example, we shall perform all further transformations
using (\ref{n9}) for the voltage $U(x,t)$. Serial expansion of
$U(x,t)$ with respect to the functions $\varphi_n(x)$ gives
\begin{equation}
\label{n12} U(x,t)=\sum\limits_n\, A_n(t)\varphi_n(x).
\end{equation}
As a result, we have:
\begin{equation}
\label{n13} \frac{\partial^2 A_n}{\partial t^2} + \omega^2_n(t)A_n
+ b(t)\frac{\partial A_n}{\partial t} + f(t) A_n=0,
\end{equation}
where $\omega_n=k_nv(t)$,

\begin{equation}
\label{n14} b(t)= v^2(t)\left[\frac{\partial L_0 C_0}{\partial t}+
L_0\frac{\partial C_0}{\partial t}+L_0g_0+R_0C_0\right],
\end{equation}

\begin{equation}
\label{n15} f(t)= v^2(t)\left[\frac{\partial }{\partial
t}\left(L_0\frac{\partial C_0}{\partial t}\right) +\frac{\partial
L_0g_0 }{\partial t}  +R_0 \frac{\partial C_0}{\partial t} + g_0
R_0\right].
\end{equation}

Thus, we get the equation of the form:
\begin{equation}
\label{n16} \frac{\partial A_n}{\partial t^2}+ b(t) \frac{\partial
A_n}{\partial t}+\Omega^2_n(t) A_n=0,
\end{equation}
where $\Omega^2_n(t)=\omega^2_n (t) +f(t)$.

If the characteristic period of the changes in the system
dimensions is long compared to the oscillation period in the
coaxial resonator, then we can seek for the solution of
(\ref{n16}) in the form

\begin{equation}
\label{n17} A_n(t)=P_{1n}(t)e^{i\int\limits_0^t\Omega_n(t')dt'}+
P_{2n}(t)e^{-i\int\limits_0^t\Omega_n(t')dt'}.
\end{equation}

Let us suppose further that $v^2=\frac{1}{L_0C_0}$ is
time-independent and the losses are absent  $(g_0=R_0=0)$.

In this case
\[
\Omega_n(t)= \sqrt{\omega^2_n (t) +
\frac{1}{L_0C_0}\frac{\partial}{\partial t}\left(L_0\frac{\partial
C_0}{\partial t}\right)},
\]
i.e.,
\begin{equation}
\label{n20} \Omega_n(t)= \sqrt{\omega^2_n (t) + \frac{1}{L_0C_0}
\frac{\partial L_0}{\partial t}\cdot\frac{\partial C_0}{\partial
t}+\frac{1}{C_0} \frac{\partial^2 C_0}{\partial t^2}}
\end{equation}
For an ordinary coaxial waveguide,  $L_0C_0=\frac{1}{v^2}=$
{const}, and so if the waveguide is contracted (or expanded), then
$L_0$ drops (increases), whereas $C_0$ increases (drops). As a
consequence, the second term in (\ref{n20}) becomes negative, and
the situation occurs when $\Omega_n(t)$ becomes imaginary, the
oscillations at the frequency $\Omega_n$ vanish, and one of the
exponents in (\ref{n17}) starts rising. A similar situation
emerges for the current in the waveguide. With reducing
longitudinal dimensions of the coaxial resonator, the frequency
$\omega_n$ builds up due to increasing  wavenumber $k_n$. As a
result, the change in geometrical dimensions of a coaxial
resonator leads to a significant change in the radiation loss in
the system.

Here we discussed a coaxial resonator (waveguide). As is known, to
describe different types of resonators in
  a stationary case,  we can  introduce inductances and
  capacitances. In a nonstationary case, we can introduce
  time-dependent $L(t)$ and $C(t)$ and
  derive equations similar to those obtained in this paper.

\end{document}